\documentstyle[12pt]{article}
\def\v{\begingroup\obeyspaces\u}
\def\u#1{\verb!#1!\endgroup}
\def\PD{{\small PDG}}
\def\HW{{\small HERWIG}}
\def\IS{{\small ISAJET}}
\def\IW{{\small ISAWIG}}
\def\SY{{\small SUSY}}
\def\qbar{\bar q}   
\begin{document}
\begin{center}{\large\bf HERWIG 6.1 Release Note}\end{center}
    A new version of the Monte Carlo program \HW\ (version 6.1) is now
    available, and can be obtained from the following web site:
\small\begin{quote}\tt
            http://hepwww.rl.ac.uk/theory/seymour/herwig/
\end{quote}\normalsize
    This will temporarily be mirrored at CERN for the next few weeks:
\small\begin{quote}\tt
            http://home.cern.ch/~seymour/herwig/
\end{quote}\normalsize
    More complete information on \HW\  can be found in the publication
    G. Marchesini, B.R. Webber,  G. Abbiendi, I.G. Knowles, M.H. Seymour
    and L. Stanco, Computer Phys. Commun. 67 (1992) 465  and also in the
    documentation for the previous version (5.9), which are available at
    the  same site, together  with other  useful files  and information.
    Here we merely give the new features relative to 5.9.

    If you use HERWIG, please refer to it something along the lines of:
\small\begin{quote}\tt
    HERWIG 6.1, hep-ph/9912396; G. Marchesini, B.R. Webber, G. Abbiendi,
    I.G. Knowles, M.H. Seymour and L. Stanco,
    Computer Phys. Commun. 67 (1992) 465.
\end{quote}\normalsize

\begin{center}{\bf NEW FEATURES OF THIS VERSION}\end{center}

  [N.B. Default values for input variables shown in square brackets.]
\begin{itemize}

  \item The main  new  features  are:  supersymmetric processes (both
     R-parity conserving and violating) in hadron-hadron collisions;
     new $e^+e^- \to$ four jets process;  matrix element corrections to
     top decay and Drell-Yan processes;  new soft underlying event
     options; updates to default particle data tables; new \LaTeX\ and
     html printout options.

  \item All R-parity  conserving \SY\ two-to-two  processes in hadron-hadron
    collisions have been added.  Their process numbers are:
\small\begin{verbatim}
    +-------+----------------------------------------------------------+
    | IPROC | Process                                                  |
    +-------+----------------------------------------------------------+
    | 3000  | 2 parton to 2 sparticles: the sum of 3010,3020 and 3030  |
    | 3010  | 2 parton to 2 spartons                                   |
    | 3020  | 2 parton to 2 gauginos                                   |
    | 3030  | 2 parton to 2 sleptons                                   |
    +-------+----------------------------------------------------------+
\end{verbatim}\normalsize
    Further details  of the inclusion of superpartners  and their decays
    are given below.

    Additional  processes  for the  \SY\  two  Higgs  doublet model  are
    currently under test and will be released shortly.

  \item All  R-parity  violating  \SY\  two-to-two  processes  via  resonant
    sleptons and squarks in hadron collisions have been added.  Their
    process numbers are:
\small\begin{verbatim}
    +-------+----------------------------------------------------------+
    | IPROC | Processes derived from the LQD term in the superpotential|
    +-------+----------------------------------------------------------+
    | 4000  | The sum of 4010,4020,4040 and 4050                       |
    | 4010  | Neutralino lepton production (all neutralinos)           |
    | 401i  | As 4010 but only the ith neutralino                      |
    | 4020  | Chargino lepton production (all charginos)               |
    | 402i  | As 4020 but only the ith chargino                        |
    | 4040  | Slepton W/Z   production                                 |
    | 4050  | Slepton Higgs production                                 |
    +-------+----------------------------------------------------------+
    | 4060  | Sum of 4070 and 4080                                     |
    | 4070  | quark-antiquark production      via LQD                  |
    | 4080  | lepton production               via LLE and LQD          |
    +=======+==========================================================+
    | IPROC | Processes derived from the UDD term in the superpotential|
    +-------+----------------------------------------------------------+
    | 4100  | The sum of 4110, 4120, 4130, 4140 and 4150               |
    | 4110  | Neutralino quark production (all neutralinos)            |
    | 411i  | As 4110 but only the ith neutralino                      |
    | 4120  | Chargino quark production (all charginos)                |
    | 412i  | As 4120 but only the ith chargino                        |
    | 4130  | Gluino quark production                                  |
    | 4140  | Squark W/Z   production                                  |
    | 4150  | Squark Higgs production                                  |
    +-------+----------------------------------------------------------+
    | 4160  | quark-quark production                                   |
    +-------+----------------------------------------------------------+
\end{verbatim}\normalsize
    In addition  the R-parity violating  decays of all  superpartners is
    included.

  \item A new process describing electron-positron annihilation to four jets
    has been added. This has \v{IPROC=600+IQ}, where a non-zero value for \v{IQ}
    guarantees production of quark flavour \v{IQ} whilst \v{IQ=0} corresponds to
    the natural flavour mix.  \v{IPROC=650+IQ} is as above but without those
    terms in the matrix element which orient the event w.r.t.\ the lepton
    beam direction. The matrix elements are based on those of Ellis Ross
    \& Terrano with orientation terms from Catani \& Seymour. The soft and
    collinear divergences are avoided  by imposing a minimum $y_{cut}$,
    \v{Y4JT}
    [.01], on the initial 4 partons. The interjet distance is calculated
    using either the Durham or JADE metrics.  This choice is governed by
    the logical variable \v{DURHAM} [\v{.TRUE.}]. Note that parameterizations of
    the volume of four-body phase  space are used: these are accurate up
    to a  few percent for $y_{cut}$  values less than 0.14.  Note, also that
    the phase space is for massless partons, as are the matrix elements,
    though a mass threshold cut is applied. Finally, the matrix elements
    for the $q\qbar gg$ \& $q\qbar q\qbar$ (same flavour quark) final states
    receive contributions from 2 colour flows each, the treatment of the
    interference terms being controlled by the array \v{IOP4JT}:
\small\begin{verbatim}
        q-qbar-g-g case:
        IOP4JT(1)=0 neglect, =1 extreme 2341; =2 extreme 3421 [0]

        q-qbar-q-qbar (identical quark flavour) case:
        IOP4JT(2)=0 neglect, =1 extreme 4123; =2 extreme 2143 [0]
\end{verbatim}\normalsize
    The scale \v{EMSCA} for the  parton showers is set equal to
    $\sqrt{s y_{min}}$
    where $y_{min}$ is the least  distance, according to the selected metric,
    between any two partons.

  \item Matrix element corrections
    to the simulation of top quark decays and Drell-Yan
    processes are now available using the same general method as already
    implemented for $e^+e^-$  annihilation and DIS\@.
    If \v{HARDME} [\v{.TRUE.}] then
    fill the missing phase-space (`dead zone') using the exact 1st-order
    M.E. result (`hard corrections').  If  \v{SOFTME}  [\v{.TRUE.}] then
    correct
    emissions in the already-populated  region of phase space  using the
    exact  amplitude for every  emission  that is  capable  of being the
    hardest so far (`soft corrections').
    \begin{itemize}
    \item For $t \to bW$ decays the routine \v{HWBTOP}
      implements hard corrections.
      \v{HWBRAN} has been modified to  implement the soft corrections to top
      decays. Since the dead zone  includes part of the soft singularity
      a cutoff is required: only gluons with energy above \v{GCUTME} [2 GeV]
      (in the top rest frame) are corrected. Physical quantities are not
      strongly dependent on \v{GCUTME} in the range 1 to 5 GeV.  For details
      see
      G.~Corcella and M.H. Seymour, Phys.\ Lett.\ B442 (1998) 417.
    \item
      For the  Drell-Yan process the routine  \v{HWBDYP} implements the hard
      corrections whilst \v{HWSBRN} has been modified to implement the soft
      corrections to the initial state radiation. For details see
      G. Corcella and M.H. Seymour, {\tt hep-ph/9908338}.
    \end{itemize}

  \item The  parameters of  the model  used  for soft  interactions are  now
    available to  the user for modification.  The model is  based on the
    minimum-bias event generator of  the UA5 Collaboration, which starts
    from a parametrization of  the $\bar{p}p$ inelastic charged multiplicity
    distribution as  a negative binomial. The parameters  are as follows
    (default  parameter  values  are  the  UA5  ones  used  in  previous
    versions):
\begin{center}
\small
\begin{tabular}{|c|l|r|}
\hline
             Name   &     Description         &Default\\
\hline
& & \\
        \v{PMBN1}  & $a$ in $\bar n =as^b+c$  & 9.11  \\
        \v{PMBN2}  & $b$ in $\bar n =as^b+c$  & 0.115 \\
        \v{PMBN3}  & $c$ in $\bar n =as^b+c$  & --9.50  \\
& & \\
        \v{PMBK1}  & $a$ in $1/k =a\ln s+b$  & 0.029  \\
        \v{PMBK2}  & $b$ in $1/k =a\ln s+b$  & --0.104 \\
& & \\
        \v{PMBM1}  & $a$ in $(M-m_1-m_2-a)e^{-bM}$ & 0.4  \\
        \v{PMBM2}  & $b$ in $(M-m_1-m_2-a)e^{-bM}$ & 2.0 \\
& & \\
        \v{PMBP1}  & $p_t$ slope for $d,u$ & 5.2  \\
        \v{PMBP2}  & $p_t$ slope for $s,c$ & 3.0  \\
        \v{PMBP3}  & $p_t$ slope for $qq$  & 5.2  \\
& & \\
\hline
\end{tabular}
\end{center}

The first three parametrize the mean charged multiplicity at
c.m.\ energy $\sqrt{s}$ as indicated. The next two specify the
parameter $k$ in the negative binomial charged multiplicity
distribution. The parameters \v{PMBM1} and
\v{PMBM2} describe the distribution of cluster masses $M$ in
the soft collision. These soft clusters are generated with a flat
rapidity distribution with gaussian shoulders. The transverse
momentum distribution of soft clusters has the form
$$
P(p_t)\propto p_t\exp\left(-b\sqrt{p_t^2+M^2}\right)
$$
where the slope parameter $b$ depends as indicated on the
flavour of the quark or diquark pair created when the
cluster was produced.

As an option, for underlying events the value of $\sqrt{s}$ used to choose
the multiplicity $n$ may be enhanced by a parameter \v{ENSOF} to allow
for an enhanced underlying activity in hard events. The actual
charged multiplicity is then taken to be $n$ plus the sum of the
moduli of the charges of the colliding hadrons or clusters.

  \item There have been a number of additions/changes to the default hadrons
    included via \v{HWUDAT}.  Here the identification of hadrons follows the
    \PD\ ('98 edition) table 13.2 with numbering according to section 31.

    New isoscalars states have been added to try to complete the $1^3D_3$,
    $1^1D_2$ and $1^3D_1$ multiplets:

\small\begin{verbatim}
           IDHW  RNAME        IDPDG      IDHW  RNAME        IDPDG
           ----  -----        -----      ----  -----        -----
            395  OMEGA_3        227       396  PHI_3          337
            397  ETA_2(L)     10225       398  ETA_2(H)     10335
            399  OMEGA(H)     30223
\end{verbatim}\normalsize
    Also the following states have been re-identified/replaced:
\small\begin{verbatim}
           IDHW  RNAME        IDPDG      IDHW  RNAME        IDPDG
           ----  -----        -----      ----  -----        -----
             57  FH_1         20333
            293  F0P0       9010221       294  FH_00        10221
             62  A_0(H)0      10111       290  A_00       9000111
             63  A_0(H)+      10211       291  A_0+       9000211
             64  A_0(H)-     -10211       292  A_0-      -9000211
\end{verbatim}\normalsize
    The $f_1(1420)$ state completely  replaces the $f_1(1520)$ in the $1^3P_0$
    multiplet, taking over 57. The $f_0(1370)$ (294) replaces the $f_0(980)$
    (293) in the $1^3P_0$ multiplet; the  latter is retained as it appears
    in the decays of several other states.  The new $a_0(1450$) states (62
    -64) replace the three old $a_0(980)$ states (290~-- 292) in the $1^3P_0$
    multiplet; the latter are kept as they appear in $f_1(1285)$ decays.

    By default production of the $f_0(980)$ and $a_0(980)$ states in cluster
    decays is vetoed.

    Also, the \PD\ numbers for the remnant particles have been changed to
    98 for \v{REMG} and 99 for \v{REMN}.

  \item Since version 6.1 contains a large number of supersymmetry processes
    several new particles have been added.

    Extra scalar bosons for the two Higgs Doublet (\SY) scenario:
\small\begin{verbatim}
           IDHW  RNAME        IDPDG      IDHW  RNAME        IDPDG
           ----  -----        -----      ----  -----        -----
            203  HIGGSL0         26       206  HIGGS+          37
            204  HIGGSH0         35       207  HIGGS-         -37
            205  HIGGSA0         36
\end{verbatim}\normalsize
    Note that the lighter neutral scalar (203) is given the non-standard
    \PD\ number 26, in order to distinguish it from the minimal SM Higgs,
    \PD\ number 25.

    Extra sfermions and gauginos for \SY\ scenarios:
\small\begin{verbatim}
           IDHW  RNAME        IDPDG      IDHW  RNAME        IDPDG
           ----  -----        -----      ----  -----        -----
            401  SSDL       1000001       413  SSDR       2000001
              |  |          |               |  |          |
            406  SST1       1000006       418  SST2       2000006
            407  SSDLBR    -1000001       419  SSDRBR    -2000001
              |  |          |               |  |          |
            412  SST1BR    -1000006       424  SST2BR    -2000006

            425  SSEL-      1000011       437  SSER-      2000011
              |  |          |               |  |          |
            430  SSNUTL     1000016       442  SSNUTR     2000016
            431  SSEL+     -1000011       443  SSER+     -2000011
              |  |          |               |  |          |
            436  SSNUTLBR  -1000016       448  SSNUTRBR  -2000016

            449  GLUINO     1000021       454  CHGINO1+   1000024
            450  NTLINO1    1000022       455  CHGINO2+   1000037
            451  NTLINO2    1000023       456  CHGINO1   -1000024
            452  NTLINO3    1000025       457  CHGINO2   -1000037
            453  NTLINO4    1000035       458  GRAVTINO   1000039
\end{verbatim}\normalsize

    The implementation of \SY\ is discussed more fully below.  Note that
    the default masses of the \SY\ particles are zero and that they have
    no decay modes. Before a \SY\ process can be simulated you must load
    the appropriate  masses and decay modes generated  using \IW\ (see
    below) or an equivalent program.

    These new states don't interfere  with the user's ability to add new
    particles as previously described.

  \item It is now possible to create particle property and event listings in
    any combination of 3 formats~-- standard ASCII, \LaTeX\ or html.  These
    options are controlled by the new, logical variables \v{PRNDEF}
    [\v{.TRUE.}]
    \v{PRNTEX} [\v{.FALSE.}] and \v{PRNWEB} [\v{.FALSE.}].  The ASCII output
is directed
    to stout (screen/log file) as in previous versions. When a listing
    of particle properties is requested (\v{IPRINT.GE.2} or \v{HWUDPR} is called
    explicitly) then the following files are produced:
\small\begin{verbatim}
      If (PRNTEX):   HW_decays.tex
      If (PRNWEB):   HW_decays/index.html
                              /PART0000001.html etc.
\end{verbatim}\normalsize
    The \v{HW\_decays.tex} file is written to the working directory whilst the
    many \v{**.html} files appear in the sub-directory \v{HW\_decays/} which must
    have been created previously.  Paper sizes and offsets for the \LaTeX\ 
    output are stored at the top of the block data file \v{HWUDAT}: they may
    need modifying to suit a particular printer. When event listings are
    requested (\v{NEVHEP.LE.MAXPR} or \v{HWUEPR} is called explicitly) the
    following files are created in the current working directory:
\small\begin{verbatim}
      If (PRNTEX):   HWEV_*******.tex       where *******=0000001 etc.
      If (PRNWEB):   HWEV_*******.html      is the event number
\end{verbatim}\normalsize
    Note the html file automatically makes links to the \v{index.html} file
    of particle properties assumed to be in the \v{HW\_decays} sub-directory.

    A new integer variable \v{NPRFMT} [1] has been introduced to control how
    many significant figures  are shown in each of  the 3 event outputs.
    Basically \v{NPRFMT=1} gives short compact outputs whilst \v{NPRFMT=2} gives
    long formats.

    Note that all the \LaTeX\ files use the package \v{longtable.sty} to format
    the tables.  Also if \v{NPRFMT=2}  or \v{PRVTX=.TRUE.} then the \LaTeX\ files
    are designed to be printed in landscape mode.

  \item There were  previously some  inconsistencies and ambiguities  in our
    conventions for the  mixing of flavour `octet' and `singlet' mesons.
    They are now:
\small\begin{verbatim}
           Multiplet   Octet         Singlet          Mixing Angle
           ---------   -----         -------          ------------
            1^1S_0     eta           eta'             ETAMIX=-23.
            1^3S_1     phi           omega            PHIMIX=+36.
            1^1P_1     h_1(1380)     h_1(1170)        H1MIX =ANGLE
            1^3P_0     MISSING       f_0(1370)        F0MIX =ANGLE
            1^3P_1     f_1(1420)     f_1(1285)        F1MIX =ANGLE
            1^3P_2     f'_2          f_2              F2MIX =+26.
            1^1D_2     eta_2(1645)   eta_2(1870)      ET2MIX=ANGLE
            1^3D_1     MISSING       omega(1600)      OMHMIX=ANGLE
            1^3D_3     phi_3         omega_3          PH3MIX=+28.
\end{verbatim}\normalsize
    After mixing the  quark content of the physical  states is given, in
    terms of the mixing angle, \v{theta}, by:
\small\begin{verbatim}
                  (ddbar+uubar)/sqrt(2)           ssbar
                  ---------------------           -----
         Octet:    cos(theta+theta_0)      -sin(theta+theta_0)
       Singlet:    sin(theta+theta_0)       cos(theta+theta_0)
\end{verbatim}\normalsize
    where  \v{theta\_0=ATAN(SQRT(2))}.   Hence, using  the  default value  of
\small\begin{verbatim}
    ANGLE=ATAN(1/SQRT(2))*180/ACOS(-ONE)
\end{verbatim}\normalsize
 for \v{theta} gives ideal mixing,
    that      is,     the      `octet'      state$=s\bar s$     and      the
    `singlet'$=(d\bar d+u\bar u)/\sqrt 2$.  This choice  is important to avoid
    large  isospin violations  in the  $1^3P_0$ and  $1^3D_1$  multiplets in
    which the octet member is unknown.

  \item A new treatment of the  colour interference terms in matrix elements
    has been introduced in this version. A non-planar, interference term
    is now shared between the planar terms corresponding to well defined
    colour flows in proportion to the size of the planar terms. Existing
    two-to-two QCD processes which have been affected are:
\small\begin{verbatim}
              Light Quarks                          Heavy Quarks
              ============                          ============
           Process            IHPRO              Process           IHPRO
           -------            -----              -------           -----
    q   +q    --> q   +q       1,2       Q   +g    --> Q   +g      10,11
    q   +qbar --> q   +qbar    5,6       Qbar+g    --> Qbar+g      21,22
    qbar+q    --> qbar+q      13,14      g   +Q    --> g   +Q      23,24
    qbar+qbar --> qbar+qbar   18,19      g   +Qbar --> g   +Qbar   25,26
                                         g   +g    --> Q   +Qbar   27,28
\end{verbatim}\normalsize
    The present and previous treatments of the interference term are the
    same for the other two-to-two QCD processes which remain unaffected.

    This new procedure has been adopted for all the \SY\ QCD processes.

    For details see: K. Odagiri, JHEP 10 (1998) 006

  \item A new process, direct $\gamma\gamma\to$ charged particle pairs has been
    added.  This has \v{IPROC=16000+IQ}: if \v{IQ=1-6} then only quark flavour
 \v{IQ}
    is produced, if \v{IQ=7,8} or 9 then only lepton flavour $e$, $\mu$ or
$\tau$ is
    produced and if \v{IQ=10} then only W pairs are produced: in these cases
    particle masses effects are included. Whilst if \v{IQ=0} the natural mix
    of quark pairs are produced  using massless MEs but including a mass
    threshold cut. The range of allowed transverse momenta is controlled
    by \v{PTMIN} \& \v{PTMAX} as usual.

  \item A new package \IW\ has been created to work with \IS\ to produce
    a file of the \SY\  particle masses, lifetimes and decay modes which
    can be read in by \HW.

    This package takes  the outputs of the \IS\ SUGRA or general MSSM
    programs and produces a data file in a format that can be read in by
    the \v{HWISSP} subroutine described below.

    In addition to the decay modes included in the \IS\ package \IW\ 
    allows for  the possibility of  violating R-parity and  includes the
    calculation of all 2-body squark and slepton, and 3-body gaugino and
    gluino R-parity violating decay modes.

  \item A new subroutine \v{HWISSP} has been  added to read the file of particle
    properties produced by the \IS\ program. In principle the user can
    produce a similar file provided that the correct format is used. The
    format should be as follows.

    First the \SY\ particle and top quark masses and lifetimes are given
    as, for example:
\small\begin{verbatim}
         65
         401 927.3980    0.74510E-13
         402 925.3307    0.74009E-13
         ....etc.
\end{verbatim}\normalsize
    That is,
\small\begin{verbatim}
         NSUSY=Number of SUSY+top particles
         IDHW, RMASS(IDHW) \& RLTIM(IDHW)
         repeated NSUSY times.
\end{verbatim}\normalsize
    Next each particle's decay modes together with their branching ratios
    and matrix element codes are given as, for example:
\small\begin{verbatim}
         6
         401  0.18842796E-01     0   450     1     0     0     0
           |               |     |     |     |     |     |     |
         401  0.32755006E-02     0   457     2     0     0     0
         6
         402  0.94147678E-02     0   450     2     0     0     0
         ....etc.
\end{verbatim}\normalsize
    That is,
\small\begin{verbatim}
         Number of decay modes for a given particle (IDK)
         IDK(*), BRFRAC(*), NME(*) \& IDKPRD(1-5,*)
         repeated for each mode.

         Repeated for each particle (NSUSY times).
\end{verbatim}\normalsize
    The order in which the decay products appear is
    important in order to obtain appropriate showering and hadronization.
    The correct orderings are indicated below.
\small\begin{verbatim}
    +----------+------------------------+------------------------------+
    | Decaying | Type of Mode           |  Order of Decay Products:    |
    | Particle |                        |   1st   |   2nd   |   3rd    |
    +----------+------------------------+---------+---------+----------+
    | Top      | 2 body to Higgs        | Higgs   | Bottom  |          |
    |          +------------------------+---------+---------+----------+
    |          | 3 body via Higgs/W     | quarks or leptons | Bottom   |
    |          |                        |    from W/Higgs   |          |
    +----------+------------------------+---------+---------+----------+
    | Gluino   | 2 body modes:          |         |         |          |
    |          | without gluon          |     any order     |          |
    |          | with    gluon          | gluon   | colour  |          |
    |          |                        |         | neutral |          |
    |          +------------------------+---------+---------+----------+
    |          | 3 body modes:          | colour  |      q or qbar     |
    |          | R-parity conserved     | neutral |                    |
    +----------+------------------------+---------+---------+----------+
    | Squark/  | 2 body modes:          |         |         |          |
    | Slepton  | Gaugino/Gluino         | Gaugino | quark   |          |
    |          | Quark/Lepton           | Gluino  | lepton  |          |
    |          +------------------------+---------+---------+----------+
    |          | 3 body modes:          |sparticle| particles from     |
    |          | Weak                   |         | W decay            |
    +----------+------------------------+---------+---------+----------+
    | Squark   | 2 body modes:          |         |         |          |
    |          | Lepton Number Violated | quark   | lepton  |          |
    |          | Baryon Number Violated | quark   | quark   |          |
    +----------+------------------------+---------+---------+----------+
    | Slepton  | 2 body modes:          |      q or qbar    |          |
    |          | Lepton Number Violated |         |         |          |
    +----------+------------------------+---------+---------+----------+
    | Higgs    | 2 body modes:          |         |         |          |
    |          | (s)quark-(s)qbar       |   (s)q or (s)qbar |          |
    |          | (s)lepton-(s)lepton    |   (s)l or (s)lbar |          |
    |          +------------------------+---------+---------+----------+
    |          | 3 body modes:          | colour  |      q or qbar     |
    |          |                        | neutral |      l or lbar     |
    +----------+------------------------+---------+---------+----------+
    | Gaugino  | 2 body modes:          |         |         |          |
    |          | squark-quark           |      q or sq      |          |
    |          | slepton-lepton         |      l or sl      |          |
    |          +------------------------+---------+---------+----------+
    |          | 3 body modes:          | colour  |      q or qbar     |
    |          | R-parity conserved     | neutral |      l or lbar     |
    +----------+------------------------+---------+---------+----------+
    | Gaugino/ | 3 body modes:          | particles in the order i,j,k |
    | Gluino   | R-parity violating     |                              |
    +----------+------------------------+---------+---------+----------+
\end{verbatim}\normalsize
    A new matrix element code has been added for these decays:
\small\begin{verbatim}
       NME = 300     3 body R-parity violating gaugino and gluino decays
\end{verbatim}\normalsize
    in addition,  an extra matrix element code has been reserved for use
    in a forthcoming version:
\small\begin{verbatim}
       NME = 200     3 body top quark via charged Higgs
\end{verbatim}\normalsize

    The indices $i,j,k$ in  R-parity violating gaugino/gluino decays refer
    to the ordering  of the indices in the  R-parity violating couplings
    in the superpotential. The convention is as in H.Dreiner, P.Richardson
    and M.H.Seymour, {\tt hep-ph/9912407}.

    Next a number of parameters derived from the \SY\ Lagrangian must be
    given. These are: the ratio of Higgs VEVs, $\tan\beta$, and the scalar
    Higgs mixing  angle, $\alpha$; the  mixing parameters for  the Higgses,
    gauginos and the sleptons; the trilinear couplings; and the Higgsino
    mass parameter $\mu$.

    Finally the  logical variable \v{RPARTY} should be  set: if  \v{FALSE} then
    R-parity is violated, and the R-parity violating couplings must also
    be supplied, otherwise not.

    Details of the \v{FORMAT} statements  employed can be found by examining
    the subroutine \v{HWISSP}.

    The integer argument in the  call to \v{HWISSP(N)} gives the unit number
    to be read from. If the data is stored in a \v{fort.N} file no further
    action is required but  if the data is to be read  from a file named
    \v{fname.dat} then appropriate \v{OPEN} and \v{CLOSE} statements must be
    added by hand:
\small\begin{verbatim}
      OPEN(UNIT=N,FORM='FORMATTED',STATUS='UNKNOWN',FILE='fname.dat')
      CALL HWISSP(N)
      CLOSE(UNIT=N)
\end{verbatim}\normalsize
    A number of sets of \SY\ parameter files, produced using \IS, for
    the standard LHC SUGRA and GMSB points are available from the \HW\
    home page:
\small\begin{quote}\tt
http://hepwww.rl.ac.uk/theory/seymour/herwig/
\end{quote}\normalsize
  \item A large number of changes have been made to enable \SY\ processes to
    be included in hadron-hadron collisions. The main changes are:
\begin{itemize}
\item The subroutine \v{HWDHQK} has been  replaced by \v{HWDHOB} which does both
      heavy quark and \SY\ particle decays.

\item The subroutines \v{HWBCON}, \v{HWCGSP} \& \v{HWCFOR} have been adapted
to handle
      the colour connections found in normal \SY\ decays.

\item The subroutine \v{HWBRCN} has been included to deal with the inter-jet
      colour connections arising in R-parity violating \SY\@. Also \v{HWCBVI}
      \v{HWCBVT} and \v{HWCBCT}  have been added to handle  the hadronization of
      baryon number violating \SY\ decays and processes. If the variable
      \v{RPARTY=.TRUE.} [default] then the old \v{HWBCON} colour connection code
      is used else the new \v{HWBRCN}
\end{itemize}
  \item The option of separate treatments for `light' and b-quark containing
    clusters are now available.   The 3 variables, \v{PSPLT} (which controls
    the mass spectrum of the fragments in heavy cluster splitting) \v{CLDIR}
    (which controls whether perturbatively produced (anti-)quarks retain
    some knowledge of their direction  in cluster decays to hadrons) and
    \v{CLSMR} (which defines to what extent the hadron and constituent quark
    directions are aligned), have been made two dimensional.
    \v{ARRAY(1)} controls clusters that do NOT contain a b quark,
    \v{ARRAY(2)} controls clusters that do     contain a b quark.
    [ Default \v{ARRAY(1)=ARRAY(2)} equivalent to earlier versions. ]

  \item A new variable \v{EFFMIN} [0.001] has been introduced, it allows the user
    to set the minimum acceptable efficiency for event generation.

  \item All hadron \& lepton masses are now given to five significant figures
    whenever possible.

  \item The treatment of the perturbative $g\to q\qbar$ vertex in the partonic
    showers  has been  improved. The  total rate  is unchanged,  but the
    angular distribution  now covers the  full range, rather  than being
    confined to the angular-ordered region as before.

  \item The treatment of the intrinsic transverse momentum of partons in an
    incoming  hadron has  been improved.   It is  now chosen  before the
    initial state  cascade is performed, and  is held fixed  even if the
    generated cascade  is rejected.  This removes  a correlation between
    the amount of  perturbative and non-perturbative transverse momentum
    generated that existed before.

  \item Space-time  positioning of clusters  is now  smeared according  to a
    Gaussian distribution of width 1/(cluster mass).

  \item For $e^+e^-$ processes with ISR a check was added requiring
   \v{TMNISR} to be  greater than the light quark threshold.

  \item The treatment of the W resonance in top decays has been improved.

  \item The common block file \v{HERWIG61.INC} has had many new variables added,
    these are listed at the top of the file.

  \item Corrections for bugs have been made affecting the following:
\begin{itemize}
 \item $\eta/\eta'$ mixing: the parameterization was nonstandard (see above).

 \item 4/5 body phase space generation: was not flat~-- affected resonance
      decays only.

 \item Drell-Yan: the overall normalization was too small by a factor 2/3
      also the t-channel contribution to $q\qbar\to q\qbar$ was incorrectly
      normalized.

 \item \v{HWHV1J}: the normalization of Z+jet  production rate was a factor 4
      too small; there was an incorrect correlation between the (signed)
      W and jet rapidities; the treatment of the W/Z Breit-Wigner lead a
      normalization error by a factor $3/\pi$.

 \item \v{HWHWPR}: there was an overall normalization error of $(M_{ff'}/M_w)^2$,
      this only affected the line shape and normalization for the $t\bar b$
      final state for which $M_{ff'}$ is large.

 \item $B_d/B_s$ mixing: an incorrect formula was used.

 \item \v{VMIN2}: the effective cut-off on the space-time distances travelled
      by light partons in a shower was incorrectly implemented. Also its
      default  value has been increased to [0.1], which affects the colour
      reconnection probability.

 \item A number of fixes to improve safety against overflowing the \v{HEPEVT}
      common block.

 \item Fix to the underlying event to prevent errors with heavy quarks.

 \item \v{HWMODK}/\v{HWIODK}: a number of corrections were made and the code
     made more robust.

 \item \v{HWURES}: the minimum threshold for the decay of diquark-antidiquark
      clusters was incorrectly set.

 \item The calculation of the top lifetime has been corrected and the QCD
      corrections  included~-- this only affects the  treatment of colour
      reconnection.

 \item The space-time positioning of clusters sometimes led to them being
      produced outside the forward lightcone.  This has been rectified.
\end{itemize}
\end{itemize}
As usual, if you wish to be removed from the \HW\ mailing list, or
    if you  know someone  who wants to  be added,  please let one  of us
    know.

    Mike Seymour,  Bryan Webber,  Ian Knowles,  Peter Richardson, Kosuke
    Odagiri, Stefano Moretti, Gennaro Corcella, Pino Marchesini

    CERN, Edinburgh, Oxford, RAL, Rochester, Milano, etc,

    16th December 1999.
\end{document}